# Immunological mechanisms and immunoregulatory strategies in intervertebral disc degeneration

*Xianyi Yan[1,†], Xi Wang[2,†], Ying Che[3,†], Peng Feng[4,\*], Ting Zhang[1,4,\*]*

1 Department of Orthopaedic, The Sixth Affiliated Hospital, School of Medicine, South China University of Technology, 528200 Foshan, Guangdong, China; lyyanxy@scut.edu.cn(X.Y.); zhangtin@pitt.edu (T.Z.)

2 Department of Trauma Orthopedics, The First Affiliated Hospital of Xinjiang Medical University, 830011 Urumqi, Xinjiang, China; candyxixi1998@163.com(X.W.)

3 School of Medicine, Shandong University of Traditional Chinese Medicine, Jinan 250355, China; cheyinglinda@163.com(Y.C.)

4 Department of Orthopaedic Surgery, University of Pittsburgh, Pittsburgh, PA 15213, USA; zhangtin@pitt.edu (T.Z.); pengfeng@pitt.edu (P.F.)

\* Correspondence: zhangtin@pitt.edu (T.Z.); pengfeng@pitt.edu (P.F.)

† These authors contributed equally to this work.

**Abstract:** Intervertebral discs are avascular and maintain immune privilege. However, during intervertebral disc degeneration (IDD), this barrier is disrupted, leading to extensive immune cell infiltration and localized inflammation. In degenerated discs, macrophages, T lymphocytes, neutrophils, and granulocytic myeloid-derived suppressor cells (G-MDSCs) are key players, exhibiting functional heterogeneity. Dysregulated activation of inflammatory pathways—including nuclear factor kappa-B (NF-κB), interleukin-17 (IL-17), and nucleotide-binding oligomerization domain-like receptor protein3 (NLRP3) inflammasome activation—drives local pro-inflammatory responses, leading to cell apoptosis and extracellular matrix (ECM) degradation. Innovative immunotherapies—including exosome-based treatments, clustered regularly interspaced short palindromic repeats/CRISPR-associated protein 9 (CRISPR/Cas9)-mediated gene editing, and chemokine-loaded hydrogel systems—have shown promise in reshaping the immunological niche of intervertebral discs. These strategies can modulate dysregulated immune responses and create a supportive environment for tissue regeneration. However, current studies have not fully elucidated the mechanisms of inflammatory memory and the immunometabolic axis, and they face challenges in balancing tissue regeneration with immune homeostasis. Future studies should employ interdisciplinary approaches—such as single-cell and spatial transcriptomics—to map a comprehensive immune atlas of IDD, elucidate intercellular crosstalk and signaling networks, and develop integrated therapies combining targeted immunomodulation with regenerative engineering, thereby facilitating the clinical translation of effective IDD treatments.

# 1. Introduction

Low back pain (LBP) is the leading cause of disability worldwide. In 2020, nearly 619 million people (≈10% of the global population) were affected, and this figure is projected to reach 843 million by 2050[1]. These trends severely impair quality of life and impose a substantial socioeconomic burden[1]. In 2019 in the United States, healthcare spending on musculoskeletal disorders totaled US $380.9 billion, with low back and neck pain accounting for US $134.5 billion[2]. IDD is the primary pathological basis of LBP. Key features include decreased disc height, nucleus pulposus (NP) dehydration, annulus fibrosus (AF) disruption, and cartilage endplate (CEP) degeneration, which lead to disc mechanical dysfunction and neurogenic pain[3,4]. Therefore, fully understanding IDD pathogenic mechanisms—and identifying novel therapeutic targets—is essential to alleviate pain and reduce the socioeconomic burden.

Recent studies highlight immune pathways as critical drivers of IDD pathogenesis[4,5]. Originating from the embryonic notochord, the intervertebral disc is shielded from systemic immunity by physical and molecular barriers—such as the AF, CEP, and additional checkpoints—making it a classical immune-privileged site[6]. The NP is highly immunogenic; intact NP tissue can strongly activate immune cells[7]. When barriers are breached by aging, mechanical overload, repetitive injury, or metabolic disturbances, the NP is exposed to immune surveillance and recognized as non-self. This exposure triggers a primary immune response[7]. Exposed NP recruits and activates mononuclear phagocytes, which release damage-associated molecular patterns (DAMPs) and trigger a surge of cytokines and chemokines[8]. This response accelerates NP degeneration, ECM degradation, and ultimately disrupts disc architecture and biomechanical function[9,10]. Furthermore, coexisting metabolic disorders can exacerbate this degenerative cascade. For example, in conditions such as obesity, diabetes, aging, and osteoporosis, elevated adipokines (e.g., leptin), increased reactive oxygen species, and metabolic dysregulation amplify inflammatory signals[11,12]. These factors skew monocytes toward a pro-inflammatory M1 phenotype, inhibit reparative M2 polarization, and create a feed-forward loop that exacerbates IDD[13]. Therefore, elucidating how immune cells are recruited and activated, mapping inflammatory signaling pathways, and identifying key immune–metabolic crosstalk nodes are crucial for advancing our understanding of IDD and developing novel therapies.

Currently, the mechanisms driving loss of immune privilege in intervertebral discs, the pathways of immune cell infiltration, the links between pro-inflammatory mediators and metabolic imbalance, and the dual roles of immune responses in repair versus degeneration remain poorly understood and actively debated. In particular, research on macrophage polarization and plasticity, the heterogeneity of T cell subsets, and the targeted modulation of inflammatory pathways has become central to disc immunology, though consensus is still lacking. This review systematically summarizes recent immunological insights into IDD pathogenesis, highlights current challenges and controversies, and evaluates immunomodulatory approaches for disc repair and regeneration, offering new perspectives and potential therapeutic targets. (Figure 1).

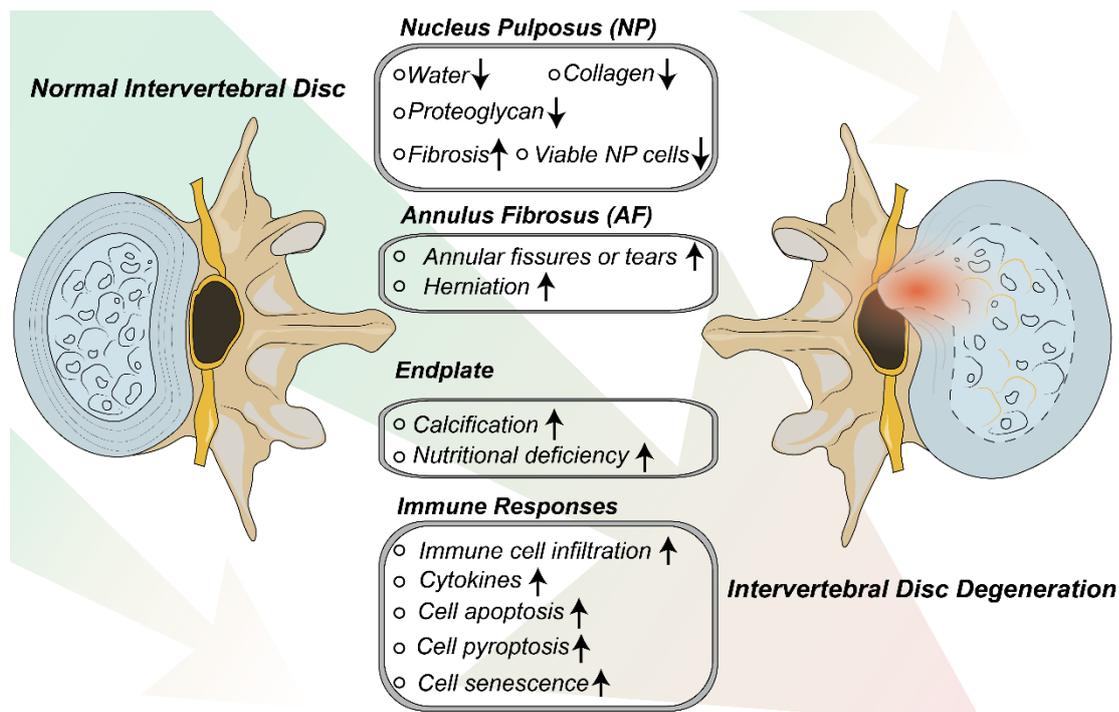

**Figure 1. Schematic of key pathological changes in IDD.**

Starting from the healthy state, discs undergo: (1) loss of NP water and proteoglycans, increased collagen deposition and fibrosis, and reduced NP cell density; (2) AF disruption with fissures leading to disc protrusion; (3) endplate calcification impairing nutrient transport; and (4) inflammatory responses—immune cell infiltration, pro-inflammatory cytokine release, apoptotic and pyroptotic cell death, and cellular senescence—which together drive IDD progression.

## 2. Participation of immune cells in the pathogenesis of IDD

IDD is a complex, multifactorial pathological process accompanied by marked biomechanical and biochemical changes[14]. Following rupture of the AF, immune cells infiltrate the disc, leading to crosstalk between the disc and the immune system[15]. Macrophages represent one of the earliest immune responders in IDD lesions, dynamically shifting between the pro-inflammatory M1 and the anti-inflammatory M2 polarization states[15]. Pro-inflammatory M1 macrophages release cytokines including tumor necrosis factor alpha(TNF-α), interleukin 1 beta(IL-1β), and interleukin 6 (IL-6), triggering activation of intracellular NF-κB[16] and mitogen-activated protein kinase (MAPK)[17] signaling cascades in NP cells, thereby exacerbating cell apoptosis and matrix catabolism. Anti-inflammatory M2 macrophages are implicated in tissue repair and fibrotic remodeling[18]. Moreover, Th1 and T helper 17 (Th17) lymphocyte subsets release interferon-γ (IFN-γ) and IL-17, respectively, potentiating local inflammation and interfacing with autoimmune pathways[19]. IFN-γ is a key cytokine secreted by Th1 cells that activates macrophages and monocytes, enhancing their phagocytic and microbicidal functions, and plays an important role in immune responses[20]. Neutrophils further intensify inflammatory insult and tissue injury through the deployment of neutrophil extracellular traps (NETs) and generation of reactive oxygen species (ROS)[21]. These intricate and persistent interactions between immune effectors and disc cells progressively amplify and derail the inflammatory milieu within the disc. Immune cells play a central role in regulating cellular senescence, and modulating immune function has emerged as a promising strategy for delaying the aging process[22]. Consequently, dissecting the stage-specific functions of immune cell subsets in IDD and mapping the regulatory networks of pivotal inflammatory pathways are instrumental for unveiling new therapeutic targets and devising innovative immunoregulatory interventions.

**2.1. Macrophage involvement in the pathogenesis of IDD**

In IDD, macrophages are pivotal mediators of pathogenesis; following disruption of the AF or CEP, they swiftly migrate into the intervertebral disc through neovessels, orchestrate the immediate release of pro-inflammatory cytokines, and regulate the local inflammatory milieu[23,24], subsequently engaging macrophage populations derived from the peripheral circulation and neighboring tissues in reparative processes of the degenerated endplates and disc matrix[25-27]. M1-polarized macrophages invading early in IDD produce abundant pro-inflammatory mediators (e.g., TNF-α, IL-1β, IL-6), which in turn trigger activation of NF-κB[28], MAPK[17], janus kinase/signal transducer and activator of the transcription (JAK/STAT)[29], toll like receptors (TLR)[30]，and NLRP3 inflammasome assembly and activation[31]，thereby initiating the inflammatory response. As the disease progresses, macrophages progressively shift toward the anti-inflammatory M2 phenotype, releasing anti-inflammatory factors such as interleukin 10 (IL-10) and transforming Growth Factor-beta (TGF-β) and activating signal Transducer and Activator of Transcription 3 (STAT3) in an effort to repair and fibrotically remodel the damaged tissue[27]. However, a prolonged chronic inflammatory state may impair normal macrophage regulatory functions, causing a severe imbalance in M1/M2 homeostasis[15]; ultimately, within this cycle of ongoing damage and repair, endplate degeneration and neovascularization are exacerbated, compromising nutrient exchange and mechanical equilibrium between the endplate and NP, and further driving irreversible disc tissue damage[15].

Notably, Luo et al. used transcriptomic analysis and protein–protein interaction (PPI) analysis to identify integrin αX (ITGAX), matrix metalloproteinase 9 (MMP9), and Fc gamma receptor IIa (FCGR2A) as hub genes predominantly expressed in macrophages and involved in regulating the inflammatory immune microenvironment of IDD[32]. Knockout of FCGR2A in macrophages inhibited M1 polarization and NF-κB phosphorylation while enhancing M2 polarization and STAT3 activation [32]. This study reveals FCGR2A as a key regulator of macrophage polarization, confirms the important role of the NF-κB/STAT3 pathway in modulating macrophage phenotype in IDD, and shows that inhibition of FCGR2A activity can significantly reduce pro-inflammatory M1 while enhancing anti-inflammatory M2, suggesting it as a potential therapeutic target for IDD[32].

Fan et al.[33] systematically investigated the role of M1 macrophage-derived exosomes (M1-Exos) in IDD. The study employed an lipopolysaccharides (LPS)-induced nucleus pulposus cell (NPC) degeneration model and a rat needle-puncture IDD model to evaluate the effects of M1-Exos on NPC senescence and disc degeneration in vitro and in vivo[33]. Results showed that M1-Exos markedly exacerbated LPS-induced NPC senescence, evidenced by increased SA-β-gal positivity, cell cycle arrest, and upregulation of senescence markers *P21* and *P53*[33]. RNA sequencing revealed enrichment of Lipocalin-2 (LCN2) in M1-Exos, which activates the NF-κB signaling pathway and thereby drives NPC senescence[33]. SiRNA-mediated silencing of LCN2 effectively suppressed expression of senescence-associated molecules and attenuated IDD progression[33]. Moreover, the exosome inhibitor GW4869 significantly reversed the pro-senescent effects of M1-Exos on NPCs, further confirming the crucial role of exosome-mediated signaling[33]. This study reveals the key mechanism by which M1 macrophages convey LCN2 via exosomes to activate NF-κB signaling, thereby promoting NPC senescence and exacerbating IDD[33].

Zhang et al.[18] investigated the mechanisms by which M2 macrophage–derived small extracellular vesicles (M2-sEVs) facilitate intervertebral disc (IVD) repair. In vitro experiments demonstrated that M2-sEVs inhibit NPC pyroptosis, maintain cell viability, and promote NPC migratory capacity[18]. Bioinformatic analyses and validation experiments showed that miR-221-3p is highly enriched in M2-sEVs and, upon transfer, downregulates expression of pyroptosis-associated factors phosphatase and tensin homolog deleted on chromosome ten (PTEN) and NLRP3, thereby inhibiting NPC pyroptosis[18]. In vivo experiments employed a decellularized ECM hydrogel encapsulating M2-sEVs for sustained release, and this system significantly delayed IDD progression[18]. This study confirms the therapeutic potential of M2-sEVs in suppressing intervertebral disc inflammation and cell death, and provides novel strategies and molecular insights for regenerative interventions in IDD[18].

## 2.2. Monocyte involvement in the pathogenesis of IDD

Monocytes serve as macrophage progenitors that swiftly home to inflammatory niches and polarize into M1 or M2 phenotypes, thereby engaging in the inflammatory cascade[15]. Within IDD pathology, dysregulated monocyte activation and continuous infiltration are deemed key contributors to the perpetuation of immune responses[34].Chen et al. identified through two-sample Mendelian randomization (MR) analysis that expression of the monocyte surface marker CD86 is significantly associated with IDD risk[35]. Mediation analysis further revealed that $CD86^+$ monocytes indirectly exacerbate IDD progression by modulating metabolites such as 3β-hydroxy-5-cholenoic acid and X-22509, with mediation effects of 9.83% and 5.11%, respectively[35]. Moreover, CD86 may exert protective effects by upregulating metabolites such as N-acetyl-L-glutamine, suggesting its bidirectional regulatory capacity under different metabolic contexts[35].

Son et al.[36] developed an innovative intervertebral disc organ-on-a-chip model to simulate monocyte chemotaxis and extravasation under degenerative nucleus pulposus conditions. The study found that degenerative NP cells release endogenous chemotactic factors that significantly enhance monocyte transendothelial migration into the NP region, independent of exogenous stimuli such as LPS or TNF-α, indicating that NP cells actively drive IDD-associated inflammation[36]. Although this study did not investigate the differentiation trajectory of monocytes in depth, the results suggest that the degenerative NP microenvironment may induce their polarization toward a pro-inflammatory macrophage phenotype, thereby contributing to the maintenance of local chronic inflammation. This work represents the first dynamic platform modeling the IDD immune microenvironment and reveals the critical role of degenerative NP tissue in monocyte recruitment and inflammatory responses, providing an experimental basis for the development of immunomodulatory strategies.

Notably, Xu et al. employed various machine learning algorithms to reveal that interferon regulatory factor 7 (IRF7) plays a significant role in the immune response of IDD[37]. Furthermore, in vivo and in vitro experiments demonstrated that knockout of IRF7 increased collagen Type II Alpha 1 Chain (COL2α1) expression while decreasing matrix Metalloproteinase 13 (MMP13), NLRP3, and IL-1β levels, attenuating IDD-associated inflammation and reversing ECM metabolic imbalance, thereby alleviating IDD progression[37]. This study reveals the pathogenic role of monocyte- and IRF7-driven mechanisms in IDD; IRF7 is a central regulator of monocyte pro-inflammatory polarization and function and, as a potential therapeutic target, provides a theoretical basis for immunoregulatory interventions in IDD.

### 2.3. T cell involvement in the pathogenesis of IDD

T cells, as central effectors of adaptive immunity, constitute an important component of the immunopathological process in IDD. Tang et al.[38] integrated multiple GEO dataset expression profiles and systematically analyzed gene expression features and immune cell infiltration in IDD using bioinformatics method. By using the CIBERSORT algorithm to assess immune cell subset infiltration, the results showed that regulatory T cells (Tregs) were significantly enriched in IDD tissues, whereas Th2 cells and dendritic cells exhibited decreased infiltration, indicating that T cell subsets have important immunoregulatory roles in IDD[38]. Further enrichment analysis showed that pathways closely related to T cell function, such as the phosphatidylinositol 3-kinase–protein kinase B(PI3K–Akt) pathway and cytokine–receptor interaction pathway, were activated in IDD, supporting the key role of T cells in the pathogenesis of IDD[38].

Notably, the transient receptor potential vanilloid 4 (TRPV4) ion channel plays a critical role in maintaining the immune privilege of healthy intervertebral discs and regulates the recruitment of T cells and monocytes in IDD[39]. Easson et al.[39] found that under mechanical loading, activation of TRPV4 triggers the NF-κB signaling pathway, resulting in increased expression of pro-inflammatory cytokines IL-6, interleukin 11(IL-11), interleukin 16 (IL-16), and leukemia inhibitory factor (LIF).

Simultaneously, TRPV4 activation also decreases the expression of T-cell chemokines—chemokine (C–C motif) ligand 3 (CCL3), chemokine (C–C motif) ligand 4 (CCL4), chemokine (C–C motif) ligand 17 (CCL17), chemokine (C–C motif) ligand 20 (CCL20), chemokine (C–C motif) ligand 22 (CCL22), and CXC chemokine ligand 10 (CXCL10)—and monocyte chemokines—chemokine (C–C motif) ligand 2 (CCL2) and chemokine (C–C motif) ligand 12 (CCL12)—potentially indirectly affecting the recruitment of these immune cells[39]. Inhibition of TRPV4 partially alleviates mechanically induced inflammatory responses and disc tissue degeneration, suggesting that TRPV4 plays a key regulatory role in interactions between the intervertebral disc and immune cells.

Clayton et al.[40] employed a mouse caudal intervertebral disc puncture model to simulate acute disc injury, collecting tissues at multiple time points and assessing immune responses via qPCR, histological staining, and flow cytometry. The study found that following IVD injury, γδ T cells (CD3$^+$CD4$^-$CD8$^-$) were significantly enriched in female mice, whereas natural killer T (NKT) cells predominated in male mice[40]. Moreover, the investigators noted temporal recruitment of macrophages and neutrophils, suggesting heterogeneous immune functions in disc repair and revealing underlying sex-specific immunological pathways[40].

Emerging evidence supports the involvement of T cells in the immune response during intervertebral disc degeneration (IDD). Future studies should further investigate the immunopathological mechanisms mediated by T cells, clarify their functional roles, and uncover novel avenues for T cell–targeted immunotherapies in IDD.

## 2.4. Neutrophil involvement in the pathogenesis of IDD

Neutrophils play an important and distinctive role in IDD. Studies have found that in IDD, infiltration of macrophages and regulatory T cells is significantly increased; CIBERSORT analysis further shows a strong positive correlation among mast cells, neutrophils, and M1 macrophages, while infiltration of B lymphocytes, regulatory T cells, and M1 macrophages is negatively correlated[34].

Zhang et al.[41] performed single-cell RNA sequencing of degenerated and normal intervertebral disc tissues, identifying a neutrophil subpopulation enriched in degenerated tissue (ECMO-neutrophils) that highly expresses macrophage migration inhibitory factor (MIF). Cell-cell communication analysis revealed that ECMO-neutrophils specifically interact with effector NPCs that highly express atypical chemokine receptor 3 (ACKR3) via the MIF/ACKR3 ligand–receptor axis[41]. Functional experiments confirmed that activation of this pathway promotes ECM degradation and the degenerative phenotype in NPCs, whereas inhibition of MIF or ACKR3 effectively alleviates these effects[41]. Analyses of patient-derived samples corroborated a direct positive association between MIF/ACKR3 expression levels and clinical IDD grade[41]. This work provides the inaugural mechanistic evidence for neutrophil-mediated IDD exacerbation via the MIF/ACKR3 axis, highlighting this pathway as a promising immunomodulatory intervention point[41].

Schweizer et al.[42] demonstrated that in Staphylococcus aureus-induced spondylodiscitis, IVD cells undergo chondroptosis—a form of programmed cell death—triggering the release of DAMPs, such as high mobility group box 1 protein (HMGB1). These molecules act as endogenous danger signals, activating the innate immune response and promoting the recruitment of neutrophils to the infected disc tissue via chemotactic factors including IL-8 and chemokine (C-X-C motif) ligand 1(CXCL1)[42]. Upon infiltration, neutrophils become activated and exert antimicrobial functions through the release of ROS, proteolytic enzymes such as elastase and myeloperoxidase, and the formation of NETs[42]. While these mechanisms aid in pathogen clearance, they simultaneously exacerbate tissue damage by degrading ECM components, including collagen and proteoglycans[42]. Consequently, neutrophil-mediated inflammation contributes to the progression of disc degeneration in the context of infection. These findings underscore the dual role of neutrophils in infectious spondylodiscitis and highlight their potential as therapeutic targets in the prevention of infection-associated disc degeneration[42].

**2.5. Involvement of additional immune cell types in IDD**

Within the IDD microenvironment, in addition to macrophages, T cells, neutrophils, and monocytes, multiple other immune cell types—including natural killer (NK) cells, B cells, and G-MDSCs—contribute to the pathological process[41]. Although these cells may not predominate numerically, they play indispensable roles in immune regulation, amplification of inflammation, and tissue injury.

As a key component of innate immunity, NK cells can induce apoptosis of target cells by releasing perforin and granzymes[43]. In IDD, NK cells may recognize stressed or damaged disc cells, induce their death, and amplify local inflammatory responses[41]. Furthermore, some studies have suggested that NK cells can cooperate with dendritic cells to enhance antigen presentation by NP cells, thereby initiating adaptive immune activation[44]. Shi et al.[45] performed single-cell RNA sequencing (scRNA-seq) to analyze the heterogeneity and cellular interactions of human degenerated intervertebral disc endplate cells. The study found intercellular interactions between T cells, NK cells, and chondrocytes; ligand–receptor pair analysis revealed that T and NK cells exert regulatory functions via molecules such as CD74 during endplate degeneration[45]. These findings indicate that NK cells may partake in the immunoregulatory processes of IDD through direct interactions with chondrocytes.

B cells primarily mediate immune regulation by presenting antigens, initiating immune responses, and producing antibodies, performing various immunomodulatory functions[46]. Rohanifar et al.[47] found that in IDD, the number of B lymphocytes increases significantly, particularly at two and eight weeks post-surgery, with marked differences compared to controls. The authors noted that B lymphocyte accumulation is a key hallmark of immune microenvironment remodeling in the intervertebral disc, suggesting that the role of adaptive immunity in IDD may be underestimated, challenging the traditional view dominated by innate immunity (e.g., macrophages)[47]. Although this study did not explore the specific functional mechanisms of B lymphocytes in depth, it is the first to reveal at the single-cell level the enrichment of B cells in IDD, indicating their potential key role in disease immunoregulation and progression, and providing new directions for future research.

Notably, G-MDSCs represent a myeloid population endowed with potent immunosuppressive activity[48]. Recent single-cell RNA-seq analyses of human discs have revealed enrichment of G-MDSCs in mildly degenerated NP tissue; these cells mitigate IL-1β-induced matrix degradation by suppressing T cell activation and ROS production, and by downregulating a disintegrin and metalloproteinase with thrombospondin motif 4 (ADAMTS-4), a disintegrin and metalloproteinase with thrombospondin motif 5 (ADAMTS-5), and matrix metalloproteinase 13 (MMP-13), thereby exerting an early protective effect[48]. Conversely, in severely degenerated disc tissue, G-MDSC frequencies are markedly reduced, suggesting that these cells play a crucial role in maintaining immune homeostasis during the early stages of IDD[48]. Moreover, G-MDSCs interact with dendritic cells and macrophages to modulate their phenotype and function, thereby indirectly shaping the immune microenvironment[48]. Therefore, G-MDSCs constitute pivotal regulators of immune balance in IDD and represent promising targets for future immunotherapeutic intervention.

In summary, the pathological progression of IDD involves the heterogeneous and coordinated activities of multiple immune cell types. NK cells, B cells, and G-MDSCs interact to regulate local inflammation, apoptosis, matrix degradation, and repair processes, collectively shaping the immune microenvironment of degenerated discs (Figure 2). Future elucidation of the molecular mechanisms and cross-regulatory interactions of these cell types will provide theoretical foundations and practical guidance for targeted immunotherapies in IDD.

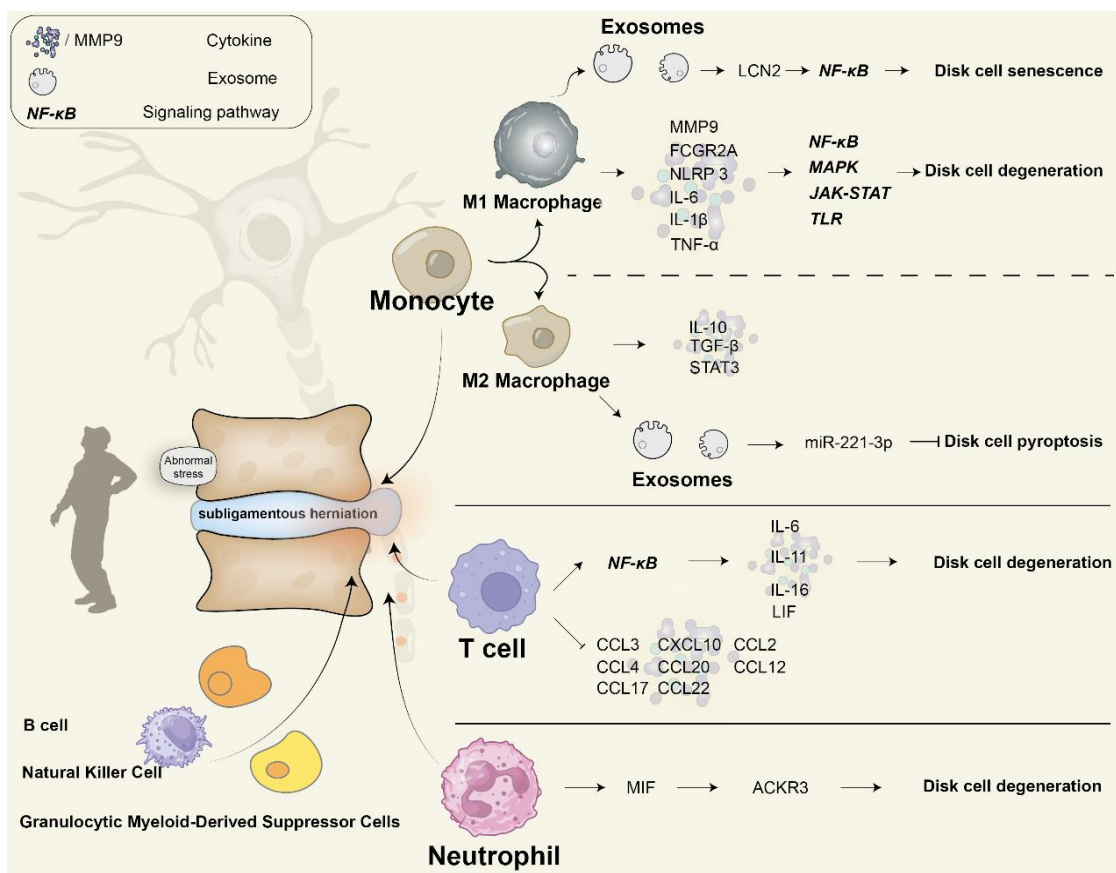

**Figure 2. Immune cell involvement in IDD.**

In degenerated intervertebral discs, monocytes, M1/M2 macrophages, T cells, neutrophils, B cells, natural killer cells, and G-MDSCs have been identified. These cells collectively participate in the initiation and progression of IDD.

## 3. Inflammatory processes in IDD

IDD is a complex, multifactorial pathological process in which inflammatory responses occupy a central role. Increasing evidence indicates that inflammatory processes promote IDD progression by modulating disc cell behaviors such as apoptosis[8], senescence[49], and ECM degradation[13]. Emerging research demonstrates that inflammatory cascades driven by NF-κB, MAPK, JAK–STAT, and TLR/NLRP3 inflammasome signaling are instrumental in the pathophysiology of IDD and protrusion[16,17,50].

### 3.1. Classic inflammatory pathways in IDD

Numerous studies have demonstrated that multiple cell types within the intervertebral disc—including NP cells, AF cells, and infiltrating immune cells (e.g., macrophages, T cells, neutrophils)—can release various pro-inflammatory cytokines and chemokines, such as TNF-α[51], IL-1α/β[51], IL-4[52], IL-6[53], IL-10[54], IL-17[55], IFN-γ[56], and prostaglandin E2 (PGE2)[57], these mediators regulate autophagy, apoptosis, and ECM degradation in disc cells via complex signaling pathways. Among these cytokines, TNF-α and IL-1β have been the most extensively studied[58,59].

#### 3.1.1. Inflammatory pathways mediated by TNF-α and IL-1β

TNF-α and IL-1β have been the most extensively studied mediators in the pathophysiology of IDD. In recent years, accumulating evidence has established TNF-α as a pivotal driver of IDD onset and progression, orchestrating disc metabolic imbalance, inflammatory activation, cell apoptosis, and ECM degradation[60]. IL-1β is regarded as a key inflammatory cytokine in IDD because it markedly induces the secretion of multiple pro-inflammatory mediators and activates inflammatory cascades[61]. Moreover, recent studies reveal that IL-1β not only exerts potent pro-inflammatory effects but also regulates disc cell programmed death pathways—including apoptosis, pyroptosis, and ferroptosis—through diverse mechanisms[61]. Additionally, IL-1β accelerates ECM degradation and impairs matrix synthesis by upregulating matrix metalloproteinase 3 (MMP-3) and matrix metalloproteinase 13 (MMP-13) expression, thereby undermining disc homeostasis and reparative capacity[62]. Simultaneously, TNF-α and IL-1β synergistically enhance syndecan-4 (SDC4) expression and activate matrix-degrading enzymes such as ADAMTS-5[63].

Tao et al.[64] used tumor necrosis factor receptor 1/2 (TNFR1/TNFR2) double-knockout mice and compared them with wild-type C57BL/6 male mice to investigate the roles of TNF-α and its receptors TNFR1 and TNFR2 in the development and progression of IDD. In a 21-month-old mouse model, the TNFR1/2 KO group exhibited protective effects—namely increased NP height, reduced endplate porosity, and more intact disc structure; receptor deletion also enhanced expression of ECM anabolic proteins (e.g., Aggrecan, COL2α1) while suppressing matrix-degrading enzymes (e.g., MMP13, ADAMTS5), indicating a shift of ECM homeostasis toward synthesis[64]. This study is the first to demonstrate at the whole-animal level that activation of the TNF-α/TNFR signaling axis is a key driver of IDD onset and progression, accelerating disc tissue destruction by modulating ECM degradation, apoptosis, and senescence.

Vera Francisco et al.[65] investigated the molecular characteristics of normal and degenerated human intervertebral disc cells using metabolomic profiling and gene expression analysis. The study found that expression of inflammation-related genes—nitric oxide synthase 2 (NOS2), cyclooxygenase-2 (COX2), IL-6, interleukin 8(IL-8), IL-1β, and TNF-α—was significantly upregulated in degenerated disc cells, indicating an important role for TNF-α in disc degeneration[65].

Zhong et al.[66] successfully established an IDD model in rats by local administration of IL-1β; MRI and histological assessments (H&E and TUNEL staining) revealed pronounced disc structural damage and cell apoptosis. In vitro, IL-1β–stimulated NPCs exhibited marked upregulation of pyroptosis-related proteins—NLRP3, caspase-1, and gasdermin D (GSDMD)—indicating that IL-1β induces NP cell pyroptosis via activation of the NLRP3 inflammasome[66]. Transcriptomic analysis further revealed that DEAD-box RNA helicase 3 (DDX3) is significantly upregulated in degenerated disc tissue. Knockdown of DDX3 effectively inhibited IL-1β–induced activation of pyroptotic signaling pathways and mitigated NP cell injury[66]. Taken together, this study elucidates the mechanism by which IL-1β promotes IDD progression through induction of NP cell pyroptosis, with DDX3 acting as a key regulatory factor[66]. Targeted inhibition of DDX3 holds promise as a potential therapeutic strategy to counteract IL-1β–mediated disc degeneration

**3.1.2. Inflammatory pathways mediated by IL-17**

Recently, the role of Th17 cells and their secreted cytokine IL-17A in IDD has received increasing attention[19,67]. Specifically, IL-17A binds TRAF6 via the SEFIR domain of nuclear factor kappa B (Act1), thereby activating TGF-β–activated kinase 1 (TAK1) and initiating the NF-κB signaling cascade[68]. Activation of this cascade induces release of pro-inflammatory mediators such as TNF-α, IL-2, IL-6, and IFN-γ, and promotes secretion of ECM-degrading enzymes (MMP-3, MMP-9, MMP-13, ADAMTS-4, ADAMTS-5), collectively driving tissue destruction in IDD[16].

Cai et al.[69] comprehensively demonstrated—using molecular RNA-seq analyses, cellular functional assays, and animal model validation—that a multiscale mechanoadaptive hydrogel delays IDD by inhibiting the IL-17 signaling pathway and improving the disc microenvironment. Specifically, IL-17A binds to its receptor IL-17R, activates the NF-κB and MAPK signaling pathways, and induces robust expression of downstream inflammatory mediators (e.g., TNF-α, IL-6) and ECM-degrading enzymes (e.g., MMPs, ADAMTS)[69]. In degenerated NPCs, IL-17A levels are significantly elevated, promoting apoptosis, autophagic dysregulation, and matrix degradation;

additionally, it enhances oxidative stress and immune cell chemotaxis, exacerbates local immune imbalance and chronic inflammation, and facilitates aberrant nerve fiber ingrowth into the NP—an important contributor to IDD-associated pain[69].

Zhou et al.[70] developed a multifunctional, microenvironment-responsive metal–phenolic network delivery platform (TMP@Alg-PBA/PVA) for treating IL-1β–induced IDD. The platform releases a metal–phenolic chelate (TA–Mn–PVP, TMP) that localizes to mitochondria, efficiently scavenges reactive oxygen species, and attenuates ECM degradation[70]. Mechanistic studies demonstrated that this system significantly inhibits pyroptosis by suppressing the IL-17/ERK signaling pathway, and in a rat IDD model it exhibited favorable therapeutic efficacy, effectively delaying disc degeneration[70].

### 3.1.3. Role of the NF-κB signaling pathway in IDD

The NF-κB signaling pathway is one of the most critical and extensively studied inflammatory regulatory pathways in IDD. This pathway can be activated by various stimuli, including TNF-α, IL-1β, oxidative stress, advanced glycation end-products (AGEs), and exosome-derived microRNAs[50]. The canonical activation route involves pro-inflammatory factors stimulating the IκB kinase complex (IKK), leading to phosphorylation and degradation of inhibitor of Nuclear Factor Kappa B Alpha (IκBα), release of the *p65/p50* heterodimer, its nuclear translocation, and initiation of transcription of inflammation-related genes[71]. During IDD, activated NF-κB induces upregulation of pro-inflammatory cytokines (e.g., IL-6, TNF-α), CCL2 and matrix-degrading enzymes (MMP-3, MMP-13, ADAMTS-4/5), resulting in ECM breakdown and disc cell apoptosis, thereby promoting degeneration[72].

Wang et al.[73] constructed a circRNA/miRNA/mRNA regulatory network using GEO datasets (GSE70362, GSE67566, and GSE116726) and identified CircEYA3 as significantly upregulated in IDD. qPCR and Western blot analyses demonstrated that CircEYA3 expression is elevated in degenerative NP cells and negatively correlates with downregulated miR-196a-5p and upregulated early B-cell factor-1(EBF1)[73]. In vitro overexpression of CircEYA3 promoted apoptosis and ECM degradation in NP cells, while inhibiting their proliferation[73]. Mechanistic investigations indicated that CircEYA3 acts as a molecular sponge for miR-196a-5p, relieving its suppression of EBF1 and thereby enhancing EBF1 expression[73]. EBF1 upregulates IKKβ, activates the NF-κB pathway, and promotes the expression of degeneration-associated genes. Animal experiments further validated the regulatory role of the CircEYA3/miR-196a-5p/EBF1 axis in IDD[73]. In summary, CircEYA3 exacerbates IDD by activating NF-κB signaling via this regulatory axis, suggesting its potential as a therapeutic target.

Liu et al.[74] found that CARMA3 expression is significantly upregulated in degenerated NP tissue, markedly promoting activation of the NF-κB signaling pathway. This process is accompanied by increased expression of multiple ECM-degrading enzymes (MMP-3, MMP-13, ADAMTS-5) and pro-apoptotic proteins (caspase-3), thereby exacerbating the onset and progression of IDD[74]. Further mechanistic studies showed that CARD-recruited membrane-associated protein 3 (CARMA3) can assemble with B cell lymphoma/leukemia 10 (BCL10) and mucosa-associated lymphoid tissue lymphoma translocation protein 1 (MALT1) to form a signalosome complex that cooperatively drives sustained activation of the NF-κB pathway. Silencing CARMA3 not only significantly inhibits formation of this complex but also effectively reduces NF-κB pathway activity, alleviating apoptosis and matrix degradation in NP cells[74].

NF-κB also synergizes with other pathways—including MAPK, PI3K/AKT, and the NLRP3 inflammasome—to maintain a chronic inflammatory microenvironment. For instance, Maltol inhibits PI3K/AKT/NF-κB signaling and NLRP3 inflammasome–driven inflammation, thereby reducing ECM degradation and inflammatory responses, upregulating anabolic protein expression, downregulating catabolic protein expression, and decreasing secretion of IL-18 and IL-1β[75]. Additionally, at the epigenetic level, NF-κB signaling is modulated by noncoding RNAs[76] and ubiquitination[77], thereby regulating inflammatory responses. For example, the membrane-associated ubiquitin ligase membrane-associated ring-CH-type finger 8 (MARCHF8) activates NF-κB by targeting transforming growth factor-beta-induced protein (TGFBI) for ubiquitin-mediated degradation, triggering inflammation and ECM breakdown, which is a key mechanism in IDD progression[78].

**3.1.4. NLRP3 inflammasome and related signaling pathways**

The NLRP3 inflammasome serves as a crucial regulatory hub for inflammatory responses in IDD; it is activated via signaling axes such as PI3K/AKT and NF-κB, leading to release of pro-inflammatory cytokines IL-1β and IL-18, which in turn exacerbate ECM degradation and disc cell apoptosis, thus establishing a vicious inflammatory cycle[8]. In recent years, an increasing number of studies have identified the NLRP3 pathway as a promising therapeutic target for IDD, and related inhibitory approaches have demonstrated potential to retard degenerative progression[79].

It is noteworthy that NLRP3 activation not only mediates classical inflammatory processes but also induces pyroptosis, a form of programmed cell death centered on GSDMD cleavage and accompanied by massive release of inflammatory cytokines[80]. Oxidative stress and mitochondrial damage–induced mitochondrial DNA (mtDNA) leakage can activate the cyclic GMP-AMP synthase-stimulator of interferon genes(cGAS–STING) pathway, which in turn triggers NLRP3 inflammasome assembly and induces pyroptosis, thereby exacerbating local immune imbalance[8]. Furthermore, NLRP3 also participates in the regulation of PANoptosis—a form of cell death that integrates pyroptosis, apoptosis, and necroptosis—further expanding its multifaceted role in the pathogenesis of IDD[81].

Mitochondrial homeostasis regulators such as sirtuin 3 (SIRT3) and PTEN induced putative kinase1(PINK1) not only play pivotal roles in controlling oxidative stress levels but also can indirectly inhibit NLRP3 inflammasome activity, thereby modulating the recruitment and activation of immune cells and shaping the immuno-inflammatory microenvironment in IDD[82,83]. Studies have shown that sustained activation of the NLRP3 pathway correlates closely with pronounced infiltration of M1 macrophages, monocytes, and T cells in degenerated NP[79]. Further experiments demonstrate that inhibition of the NLRP3 inflammasome or its downstream signaling significantly reduces ECM degradation and decreases secretion of pro-inflammatory cytokines such as IL-1β and IL-18, thereby alleviating local inflammation and delaying IDD progression[75]. In summary, the NLRP3 inflammasome represents a central regulatory node of IDD-associated inflammation and, by engaging multiple programmed cell death pathways, mediates disc cell dysfunction and tissue damage—positioning it as a promising therapeutic target in IDD immunopathology.

### 3.1.5. Inflammatory responses and inflammatory memory

Inflammatory memory refers to the phenomenon whereby tissues, after encountering an inflammatory insult, undergo mechanisms such as epigenetic reprogramming to mount enhanced responses to future stimuli, as evidenced by monocytes, macrophages, and/or natural killer cells exhibiting amplified immune reactions to microbial pathogens upon re-stimulation[84]. For instance, in systemic lupus erythematosus patients, transcriptomic reprogramming and myeloid skewing of hematopoietic stem and progenitor cells result in increased circulating neutrophil counts, heightened inflammatory activity, and accelerated disease progression[85]. In gouty arthritis, monocytes become hyperinflammatory upon exposure to soluble urate, leading to increased production of pro-inflammatory cytokines upon re-stimulation; this effect represents inflammatory memory mediated by epigenetic and metabolic reprogramming[86,87].

The initiation of inflammatory responses in IDD fundamentally arises from endogenous (e.g., metabolic dysregulation, oxidative stress) or exogenous (e.g., mechanical injury, microbial infection) stimuli to the immune system, which in turn trigger metabolic and epigenetic reprogramming[88]. As detailed above, various immune cells—including macrophages, T cells, and B cells—infiltrate degenerated disc tissue en masse, establishing a complex local inflammatory milieu[4], and the cytokines they secrete not only influence macrophage M1/M2 polarization[15] but also induce disc cell apoptosis and ECM degradation[89].

Thus, a chronic inflammatory milieu may induce an "inflammatory memory" state in local macrophages and NPCs—manifested as an exaggerated inflammatory response upon re-stimulation; although this mechanism has not yet been directly confirmed, it opens new avenues for future elucidation of IDD pathogenesis. Therefore, future investigations should center on inflammatory memory and the heterogeneity of disease pathology, examining the concomitant intracellular metabolic and epigenetic alterations under pathological conditions to optimize the therapeutic potential of targeting inflammatory memory.

### 3.2. Immune-metabolic axis and intervertebral disc inflammation

Under conditions of metabolic syndrome such as obesity or type 2 diabetes, abnormal expansion of adipose tissue disrupts systemic immune homeostasis[90]. Pro-inflammatory adipokines such as leptin and resistin are markedly elevated, whereas secretion of the anti-inflammatory adiponectin is significantly reduced, creating a pro-inflammatory immunometabolic microenvironment[91]. These adipokines bind to cognate macrophage receptors, activating inflammatory signaling pathways including NF-κB, JNK, and MAPK, thereby driving polarization toward the pro-inflammatory M1 phenotype and inhibiting anti-inflammatory M2 conversion[15]. M1 macrophages secrete abundant pro-inflammatory cytokines—TNF-α, IL-1β, and IL-6—accelerating disc cell apoptosis and ECM degradation, thereby exacerbating the degenerative process[15]. A growing body of evidence indicates that metabolic byproducts—cholesterol, fatty acids, lactate, and homocysteine—and metabolic reprogramming processes play critical roles in regulating NP cell biology[92-95]. Studies have demonstrated that atorvastatin can alleviate TNF-α–induced NP degradation through multiple mechanisms, showing potential as a disc-protective agent[96]. It not only suppresses expression of matrix-degrading enzymes (MMP-3, MMP-13, ADAMTS4) and restores collagen II and aggrecan levels, but also significantly inhibits NLRP3 inflammasome activation, reduces caspase-1 and GSDMD expression, and attenuates inflammatory responses[96]. Simultaneously, atorvastatin enhances autophagy, as evidenced by an increased ratio of lipidated microtubule-associated protein 1 light chain 3 (LC3-II) to its non-lipidated form (LC3-I) and decreased p62, thereby maintaining cellular homeostasis[96]. Further research indicates a negative regulatory relationship between NLRP3 inflammasome activation and autophagy; atorvastatin inhibits the NF-κB pathway, co-regulating inflammasome activity and autophagy to exert a multi-targeted intervention in disc degeneration[96].

Hyperglycemia and high-fat diets can induce lipid accumulation, insulin resistance, and chronic inflammation, disrupting intervertebral disc homeostasis and serving as significant risk factors for IDD[97]. Obesity and hyperglycemic conditions markedly elevate ROS levels, exacerbating oxidative stress[98]; ROS not only directly damage disc cell DNA and mitochondrial function but also amplify inflammation via NF-κB and NLRP3 inflammasome activation, establishing a vicious cycle between inflammation and metabolism[99]. Gallate et al.[100] systematically investigated the pathogenic role of AGEs in IDD and compared the differential regulatory effects of their receptors RAGE and Galectin-3 (Gal3) in this context. Studies indicate that high-sugar diets or diabetic states promote AGE accumulation in disc tissue, which by inducing oxidative stress and enhancing catabolic activity leads to collagen damage and mechanical property decline, representing a key metabolic driver of IDD[100]. Functional assays reveal that RAGE is significantly upregulated upon AGE stimulation, and its activation constitutes a key mechanism by which AGEs mediate disc collagen degradation, suggesting a pro-pathogenic role in degeneration[100]. Conversely, Gal3 expression is downregulated under AGE exposure but exerts tissue-protective effects, including reducing collagen damage and preserving the biomechanical stability of the disc[100]. This study is the first to demonstrate AGE-mediated direct tissue damage in a mouse disc organ culture model and to elucidate the functional disparities between RAGE and Gal3 in regulatory processes[100]. The findings suggest that targeted inhibition of RAGE or enhancement of Gal3 activity could serve as potential therapeutic strategies for IDD in the context of dysregulated glucose metabolism[100].

## 4. Epigenetic regulation in the pathogenesis of IDD

Accumulating evidence indicates that the immune-inflammatory processes in IDD are influenced by epigenetic modifications[101]. In late-stage degenerative intervertebral discs, within an immune-inflammatory milieu, epigenetic modifications of genes associated with the inflammatory response are aberrantly regulated during IDD, thereby promoting inflammation and disease progression[101].

### 4.1. Functions of DNA and RNA methylation in IDD

In recent years, an increasing number of studies have demonstrated that DNA and RNA methylation modifications play a crucial role in the pathological processes of IDD by regulating gene expression and signaling pathway activity[101]. Among these, DNA methylation is a widespread epigenetic modification primarily catalyzed by members of the DNA methyltransferase (DNMT) family—DNMT1, DNMT3A, and DNMT3B—and broadly influences the stability of gene expression[102]. During IDD, aberrant alterations in DNA methylation are closely associated with pathological events such as inflammatory responses, cell death, and oxidative stress[103,104].

Specifically, in severely degenerated NP tissues, studies have identified pronounced DNA hypermethylation at the promoter region of the iron transporter ferroportin, and this hypermethylation is significantly associated with elevated expression of DNA methyltransferase DNMT3B[105]. DNMT3B exacerbates IDD progression by increasing methylation at the ferroportin promoter, thereby downregulating its expression and promoting NP cell ferroptosis and oxidative stress[105]. Notably, treatment with DNA methyltransferase inhibitors such as 5-azacytidine (5-AZA) effectively reduces ferroportin promoter methylation, restores its expression, and markedly enhances disc cell viability, suggesting that targeting DNA methylation pathways may offer a promising strategy for IDD therapy[105].

Furthermore, RNA methylation modifications, particularly $N^6$-methyladenosine (m6A) methylation, have gradually been recognized as playing a significant role in the progression of IDD[106]. As a prevalent RNA epigenetic mark, m6A is a prevalent RNA epigenetic mark that is catalyzed by the methyltransferase complex comprising methyltransferase-like 3 (METTL3) and methyltransferase-like 14 (METTL14) and is removed by demethylases such as AlkB homolog 5 (ALKBH5) and fat mass and obesity-associated protein (FTO)[107]. Recent studies have shown that the RNA demethylase ALKBH5 is significantly upregulated in degenerated intervertebral disc tissues[108]. Mechanistically, upregulation of ALKBH5 inhibits m6A modification of DNMT3B transcripts, resulting in decreased recognition by YTHDF2 and thereby suppressing transcript decay[106]. Stabilized mRNA enhances DNMT3B expression, which in turn methylates the E4F1 promoter, suppressing E4F1 expression, exacerbating disc cell senescence and inflammatory responses, and promoting disc degeneration [106]. Conversely, m6A modifications may also modulate immune processes by impacting the expression of various inflammatory cytokines and apoptosis-related genes [109]. For instance, silencing of the m6A methyltransferase METTL14 activates the NF-κB signaling pathway, resulting in upregulation of inflammatory cytokines such as IL-6 and IL-8 [109]. Additionally, m6A reader proteins such as YTHDF2 selectively bind $m^6A$ -modified transcripts, accelerating the degradation of target mRNAs and thereby regulating disc cell survival and functional homeostasis [106].

Taken together, DNA and RNA methylation modifications play a pivotal role in the pathological mechanisms of IDD. DNA methylation induces ferroptosis and oxidative stress by modulating the expression of key genes such as ferroportin. RNA methylation, on the other hand, exacerbates disc inflammation and cellular senescence through regulation of DNMT3B and genes related to inflammatory pathways. These findings provide new theoretical foundations for elucidating the pathogenesis of IDD and suggest that targeting DNA and RNA methylation modifications may represent a novel epigenetic therapeutic approach to slow disc degeneration.

## 4.2. The role of histone modifications in IDD

Zeste homolog enhancer 2 (EZH2) is a key epigenetic regulator and is considered a driver of epigenetic modulation[110]. Increasing evidence indicates that EZH2 participates in the biology of degenerative musculoskeletal diseases through epigenetic regulation, including osteogenic–adipogenic differentiation of bone marrow mesenchymal stem cells, osteoclast activation, and chondrocyte functional status[111-113]. Overexpression of EZH2 suppresses miR-129-5p expression via H3K27me3 modification, upregulates MAPK1 expression, and leads to the release of inflammatory and senescence factors by NP cells, thereby promoting IDD progression[114]. Dysfunctional EZH2 can inhibit immune cell migration and enhance Treg suppressive activity to facilitate immune evasion[115]. These findings suggest that EZH2 may represent a highly promising target for future immunotherapies in IDD.

SIRT6, a member of the sirtuin family, possesses $NAD^+$-dependent protein deacetylase and ADP-ribosyltransferase activities and is primarily localized in the nucleus[116]. SIRT6 plays critical roles in regulating genomic stability, metabolism, inflammation, cellular senescence, tumorigenesis, and lifespan by deacetylating histone and non-histone proteins or ADP-ribosylating specific protein substrates[117,118]. Studies have shown that SIRT6 is recruited to promoters of NF-κB downstream target genes, where it deacetylates histone H3 lysine 9 (H3K9), reduces promoter acetylation, attenuates NF-κB signaling, and thereby decreases inflammatory cytokine expression[119].

HDAC4, a member of class IIa of the histone deacetylase (HDAC) family, plays a critical role in the inflammatory response of IDD[120]. Studies have demonstrated that upregulation of HDAC4 exacerbates IDD symptoms, whereas silencing HDAC4 alleviates these symptoms[121]. It was found that glycogen synthase kinase 3 beta (GSK3β) promotes HDAC4 degradation, resulting in downregulation of the downstream krüppel-like factor 5(KLF5)/apoptosis signal-regulating kinase 1 (ASK1) axis and thereby delaying IDD progression[121]. Additionally, upregulation of zinc transporter ZIP4 mediates the HDAC4–FoxO3a axis to promote inflammation and oxidative stress, exacerbating NP cell degeneration[122]. Specifically, in NP tissues from IDD patients, ZIP4 expression is elevated, and upon IL-1β stimulation of NP cells, this upregulation promotes inflammation, oxidative stress, and ECM degradation, whereas ZIP4 knockdown yields opposite effects[122]. Mechanistic studies revealed that ZIP4 upregulates HDAC4, augments NF-κB phosphorylation, and concurrently inhibits phosphorylation of SIRT1 and FoxO3a[122].

In summary, these studies demonstrate that epigenetic modifications play a pivotal role in the onset and progression of IDD, with the DNA methyltransferase, SIRT, and HDAC families closely linked to inflammatory responses during disc degeneration.

## 5. Latest advances in immunomodulatory strategies for IDD treatment

In recent years, with an ever-deepening understanding of the immune system's roles in inflammation and tissue regeneration, an increasing number of studies have shifted toward exploring immunomodulatory therapeutic strategies aimed at delaying or even reversing the progression of IDD[4]. Current studies indicate that the infiltration of immune cells—such as macrophages, T cells, and B cells—and the resultant local inflammatory response are critical events in the initiation and progression of IDD; modulating these responses can both improve the local microenvironment and promote disc cell survival and regeneration[7,123]. The following sections will provide a detailed exploration of the latest advances and remaining challenges in pharmacological treatments, stem cell–based immunotherapies, and exosome technologies, and will propose directions for future research.

**5.1. The role of immunomodulatory drugs in the treatment of IDD**

Immunomodulatory drugs, as important non-surgical strategies for IDD treatment, have garnered considerable attention in recent years. Immunomodulatory drugs, as important non-surgical strategies for IDD treatment, have garnered considerable attention in recent years. For example, M1 macrophages secrete abundant pro-inflammatory cytokines (such as TNF-α, IL-1β, IL-6), whereas M2 macrophages tend to secrete anti-inflammatory factors, thereby promoting repair[15]. In response to this phenomenon, preclinical and laboratory studies have attempted to use specific immunomodulators to induce M2 polarization, thereby reducing inflammation and promoting ECM remodeling [124,125]. Biologics such as monoclonal antibodies and cytokine antagonists can be employed for the treatment of disc degeneration. These agents target specific immune pathways to modulate intradiscal immune responses, alleviating inflammation and facilitating tissue repair. For instance, the anti-IL-6 receptor monoclonal antibody tocilizumab, administered epidurally to the spinal nerves of patients with lumbar spinal stenosis, can effectively reduce radicular leg pain, numbness, and low back pain without adverse effects [126]. A preclinical ex vivo study found that the TNF-α inhibitor etanercept effectively reduced expression of IL-1β, IL-6, IL-8, MMP1, and MMP3 in NP and AF tissues, neutralizing the pro-inflammatory and catabolic environment in IDD models [127]. Additionally, given that IL-17A secreted by Th17 cells can inhibit NP cell proliferation and ECM synthesis[19,55], IL-17AR antagonists may represent a novel therapeutic avenue for IDD. Furthermore, some studies have investigated multi-target intervention strategies—such as drugs that concurrently modulate NF-κB and MAPK signaling pathways—which may more effectively control inflammatory cascades [128]. In conclusion, immunomodulatory drugs show promising research prospects in the treatment of IDD.

**5.2. The role of mesenchymal stem cell immunotherapy in the treatment of IDD**

Mesenchymal stem cell (MSC) therapy has been officially recognized by the World Health Organization as a primary research focus[129], and because of their multipotent differentiation capacity and intrinsic immunomodulatory functions, MSCs have emerged as a hot topic in regenerative treatments for IDD[130-132]. Recent studies have revealed that MSCs can not only differentiate into chondrocyte-like cells to replenish degenerated tissue, but also modulate the local immune environment. For instance, MSCs secrete a range of cytokines and growth factors that suppress inflammatory responses, promote endogenous cell repair, and regulate macrophage polarization[133,134]. Current research primarily focuses on strategies to enhance MSC survival and engraftment within the degenerative disc environment, such as using hydrogels loaded with the chemokine stromal cell–derived factor-1 (SDF-1) to recruit endogenous MSCs and accelerate tissue repair[135,136]. Pereira et al. developed a hydroxyapatite (HAP) hydrogel incorporating SDF-1, which effectively recruits increased numbers of MSCs in IVD tissue, providing an optimal matrix for their survival and differentiation[136]. Cunha et al. utilized a chemotactic induction delivery system to embed MSCs in HAP hydrogels within the IVD, effectively enhancing COL-2 and MMP3 expression[137]. Moreover, future combination of MSCs with immunomodulatory agents (such as co-administration of MSCs with anti–TNF-α drugs) may exhibit synergistic effects by improving the local immune microenvironment, simultaneously suppressing inflammation and enhancing tissue regeneration, thus offering a multidimensional intervention strategy for IDD treatment[138].

The primary challenge facing MSC therapy at present is the low survival and engraftment rates of the cells within the degenerative microenvironment. Future studies should focus on optimizing MSC preconditioning methods—such as hypoxic culture, genetic modification, or co-application with microenvironment-modifying biomaterials—to enhance their survival and functional capacity. Moreover, elucidating the molecular and cellular mechanisms underlying the combined use of MSCs and immunomodulatory agents requires further in-depth investigation to establish a foundation for clinical translation.

### 5.3. The role of exosomes in the IDD

As intercellular communication vehicles, exosomes have shown promising applications in IDD research in recent years[139]，Exosomes can carry miRNAs, proteins, and other bioactive molecules to modulate target cell signaling and remodel the local inflammatory and immune microenvironment[139]. Thus, exosomes can influence ECM catabolism and anabolism by remodeling the inflammatory and immune status of the IVD microenvironment[140,141]. Hu et al.[140] found that hypoxia-preconditioned exosomal vesicles (HP-sEVs) alleviate the inflammatory microenvironment induced by IDD and promote NP cell proliferation, proteoglycan synthesis, and collagen formation. Transcriptome sequencing further revealed that HP-sEVs deliver miRNA-7-5p to inhibit the NF-κB/Cxcl2 axis, thereby promoting ECM regeneration and yielding superior therapeutic outcomes in IDD[140].

Notably, the biological functions of exosomes not only regulate the inflammatory microenvironment but also improve the immune milieu in IDD[4]. Zhao et al. found in NP tissues from IDD and idiopathic scoliosis patients that exosomes released by degenerated NP cells upregulate miR-27a-3p, which targets the peroxisome proliferator-activated receptor γ(PPARγ)/NF-κB/PI3K/AKT pathway to drive macrophage M1 polarization and accelerate disc fibrosis, whereas inhibition of miR-27a-3p can partially reverse this process[142].

Beyond elucidating molecular mechanisms, optimal exosome delivery routes, dosing regimens, and long-term safety remain key research questions. Numerous studies have confirmed that exosome-delivered miRNAs exert pronounced therapeutic effects in IDD by modulating not only the inflammatory microenvironment but also the immune milieu[143]. However, research on exosome-mediated miRNA therapies for IDD still faces limitations, including optimal dosing and route of administration (intradiscal versus systemic delivery)[144]. Systemic administration may require multiple doses to sustain efficacy, whereas direct intradiscal injection carries risks such as potential needle-induced damage to the vertebrae or disc tissues[145]. Although exosome-based therapies for IDD present multiple advantages, significant technical bottlenecks remain between in vitro studies and clinical translation. Future work should systematically investigate exosome biology, optimize manufacturing and delivery methods, and integrate animal model and preclinical data to comprehensively evaluate efficacy and safety. Furthermore, multimodal combination approaches (e.g., coadministration of exosomes and MSCs) may yield synergistic effects in modulating local inflammation and promoting regeneration. Immunomodulatory strategy–based intervention for disc degeneration (Figure 3).

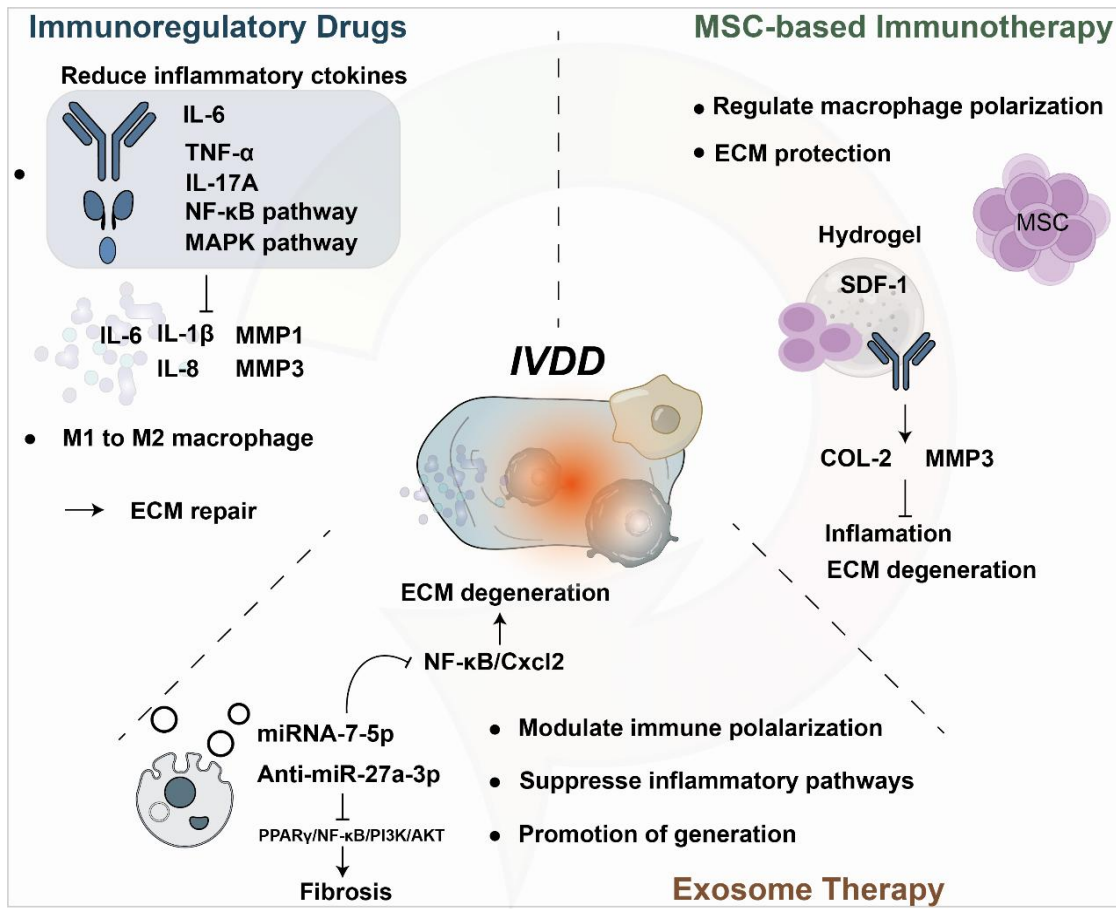

**Figure 3 IDD treatment via immunoregulatory strategies.**

Immunomodulatory therapies primarily include:(1) monoclonal antibodies and cytokine antagonists that target proteins such as IL-6, TNF-α, and IL-17, and regulate inflammation via specific signaling pathways;(2) MSC-based therapy, delivered in hydrogel matrices to reshape the immune niche and enhance tissue regeneration;(3) exosomal approaches, employing miRNA cargo to suppress inflammatory and fibrotic responses and boost regeneration.

## 6. Prospective research directions

In response to the interplay among inflammation, immunity, and metabolism within the pathological microenvironment of IDD, future research and therapeutic strategies are progressively shifting from single anti-inflammatory approaches toward an integrated "immuno-regenerative" paradigm. The following sections will articulate future research directions and potential breakthroughs across four key areas.

### 6.1. Investigating the immune–metabolism axis and inflammatory memory

An in-depth investigation into how the immunometabolic axis governs disc degeneration will be a pivotal area of study. This includes studies on mechanisms of inflammatory memory and immune cell reprogramming, with an emphasis on the interplay between systemic metabolic disorders (such as obesity and diabetes) and local inflammatory responses[146]. A growing body of evidence indicates that lipid metabolic disorders and pro-inflammatory adipokines resulting from metabolic diseases like obesity can alter disc cell phenotypes and exacerbate local inflammation[147]. A chronic inflammatory environment can also induce metabolic and epigenetic reprogramming of immune cells via pro-inflammatory cytokines, establishing an "inflammatory memory" effect that renders inflammatory responses persistently hyperresponsive[146]. Future studies should elucidate the role of such innate immune training effects in IDD, understanding how prior inflammatory or metabolic stress primes disc tissue for amplified immune responses to subsequent injury, thereby uncovering mechanisms by which systemic metabolic dysregulation and local sterile inflammation synergistically drive IDD.

**6.2. Revealing immune cell heterogeneity and communication networks**

By employing cutting-edge techniques such as single-cell RNA sequencing, spatial transcriptomics, and high-parameter immunophenotyping (e.g., mass cytometry), it is possible to comprehensively characterize the immune composition and its dynamic changes in degenerated intervertebral discs[148]. Existing single-cell RNA sequencing studies have identified multiple immune cell subsets (such as macrophages and T cells) and revealed their interaction networks with stromal cells[148]. Studies have shown that infiltrating macrophages and T cells in the degenerated NP secrete pro-inflammatory mediators such as TNF-α, inducing mesenchymal-like cells to adopt an osteogenic phenotype, thereby accelerating endplate ossification and disc degeneration; notably, anti–TNF-α agents can reverse this process[148]. Moreover, spatial transcriptomics and similar methods can localize immune cells within the disc microarchitecture and chart intercellular communication maps, complementing single-cell analyses[149].

Through multi-omics integrative analyses, key immune cell subsets and their secreted factors in IDD can be identified as potential therapeutic targets [149]. For example, studies have found that degenerated disc cells produce various chemokines (such as MCP-1 and CCL3), which attract and activate macrophages and T cells into the lesion following AF disruption [148]. This immune cell–chemokine axis not only amplifies local inflammation but also offers strategies for targeting these mediators (e.g., blockade of the MCP-1/C-C Chemokine Receptor Type 2 (CCR2) pathway) [150]. Future studies should further define the functions of pro-inflammatory M1 macrophages and Th17 cells in distinct degeneration phases, along with how deficits in anti-inflammatory M2 macrophages and Treg cells contribute to disease progression, thus pinpointing the highest-value cell subsets and signaling routes for therapeutic targeting.

### 6.3. Development of precision immunointervention strategies

Future IDD therapies should focus on precisely modulating dysregulated immune responses to interrupt the vicious cycle. For adaptive immunity, targeting hyperactive Th17 cells while enhancing regulatory T cell function may restore inflammatory balance[151]. In innate immunity, regulating macrophage polarization to shift from pro-inflammatory M1 to reparative M2 phenotypes can alleviate chronic local inflammation in the disc[15]. Studies have shown that exosomes released by degenerated disc cells carrying molecules such as miR-27a-3p induce macrophage polarization toward M1, exacerbating tissue inflammation and degeneration; blocking this exosomal signal may mitigate the inflammatory cascade and slow IDD progression[142].

Additionally, the NLRP3 inflammasome, as a pivotal trigger of sterile inflammation, is critically implicated in disc degeneration when hyperactivated and has emerged as a promising therapeutic target. Inhibition of the NLRP3 pathway (for example, targeting upstream PDE4B or directly blocking NLRP3 activity) significantly delays disc degeneration progression in model systems[152]. With advances in biotechnology, novel approaches can endow immune modulation with greater precision and feasibility. For instance, CRISPR/Cas9-mediated gene editing or activation techniques can upregulate anti-inflammatory or regenerative genes to improve the disc microenvironment at its root[153]. Studies have applied CRISPR activation in mesenchymal stem cells to enhance endogenous levels of the anti-inflammatory factor TNF-stimulated gene 6 protein (TSG-6), thereby potentiating the regenerative and immunomodulatory functions of MSC-derived exosomes[153]. Meanwhile, novel delivery platforms such as nanocarriers offer targeted transport of immunotherapeutics, enabling anti-inflammatory factors to more efficiently reach disc lesions[154]. Smart biomaterials like chemokine-laden hydrogels, serving as injectable scaffolds, can locally release drugs or cytokines and dynamically adjust release kinetics in response to inflammatory stimuli, thereby finely remodeling the disc microenvironment[136]. The integration of these strategies will markedly enhance the specificity and efficacy of immunointerventions and accelerate translation to the clinic.

**6.4. Integrating immuno-regenerative strategies with tissue engineering**

A key future direction in IDD therapy is to develop an integrated treatment paradigm that combines "immuno-regeneration" with tissue engineering, simultaneously suppressing harmful inflammation while activating and enhancing the body's regenerative capacity to achieve synergistic immune modulation and structural repair. Recent studies have shown that improving the local immune microenvironment often creates favorable conditions for tissue regeneration. For example, exosomes derived from MSCs can not only reduce the release of pro-inflammatory cytokines and degradative enzymes by inhibiting NF-κB–mediated inflammatory cascades, but also promote disc cell survival and proliferation, thereby optimizing ECM reconstruction[155], This "multifunctional" immunomodulatory effect provides a solid theoretical foundation for immuno-regenerative strategies.

Meanwhile, the advent of novel biomaterials has provided new technological avenues for integrating immunomodulation with tissue engineering. Smart hydrogels[156], injectable scaffolds[157], and nanocarrier drug delivery systems[158] not only serve as vehicles for cells or therapeutics, enabling localized sustained release of modulatory factors, but also dynamically adjust release rates in response to inflammatory cues, thereby providing mechanical support and a physiologically adaptive environment for tissue regeneration while preserving immune homeostasis. Multidisciplinary combinatorial intervention strategies are expected to overcome the limitations of conventional single anti-inflammatory or cell transplantation approaches by reconstructing a regenerative-friendly immune microenvironment, thereby further restoring the structure and function of degenerated intervertebral discs.

In summary, future IDD therapies will trend toward combining precision immunomodulation with tissue engineering techniques, forming an integrated "immuno-regenerative" paradigm from fundamental mechanisms through to clinical translation. By integrating immunotherapies, gene editing, smart delivery systems, and biomaterial technologies, entirely novel, multidimensional intervention strategies may be developed for the repair and functional restoration of degenerated discs, thereby overcoming current clinical bottlenecks and enabling personalized precision treatments.

## 7. Conclusions

IDD is a multifactorial, complex, and dynamic pathological process whose progression is regulated by local inflammation, immune cell infiltration, and metabolic dysregulation. Recent studies have demonstrated that immune cells such as macrophages, T cells, neutrophils, and G-MDSCs play crucial roles in IDD by secreting pro-inflammatory cytokines and degradative enzymes that drive apoptosis and ECM degradation. Signaling pathways including NF-κB, IL-17, NLRP3, MAPK, and TLR play central roles in regulating the inflammatory response, providing the molecular underpinnings of disease progression. Meanwhile, emerging mechanisms—such as immunometabolism, inflammatory memory, and immune cell reprogramming—offer novel perspectives for a deeper understanding of IDD.

In the future, constructing a high-resolution immunological atlas of IDD using cutting-edge approaches such as single-cell, multi-omics, and spatial transcriptomics will aid in pinpointing key pathogenic cell populations and molecular targets, thereby advancing precision immunointervention strategies. Concurrently, an "immuno-regenerative" strategy that integrates gene editing, intelligent delivery systems, and tissue engineering holds promise for achieving synergistic control of inflammation and tissue repair, thereby transforming the conventional IDD treatment paradigm. In summary, a deep integration of immunology and regenerative medicine research is anticipated to provide novel theoretical foundations and clinical translation pathways for the prevention and treatment of disc degeneration, ultimately improving patient quality of life.

**Author Contributions:** T.Z. and P.F. conceptualized the study. X.Y., T.Z., X.W.,Y.C., and P.F. contributed to writing the manuscript. X.Y., T.Z., Y.C.,and X.W. drafted the original version. P.F. prepared the illustrations using BioRender. T.Z. and P.F. reviewed, edited, and supervised the work. All four authors have participated sufficiently in this work and agree to be accountable for all aspects of the manuscript. All authors have read and approved the final version for publication.

**Funding:** This research received no external funding.

**Institutional Review Board Statement:** Not applicable.

**Data Availability Statement:** Not applicable.

**Acknowledgments:** Figures were created using BioRender.com and Adobe

Illustrator.

**Conflicts of Interest:** The authors declare that they have no conflict of interest.

## Abbreviations

| Abbreviation | Full Form |
|---|---|
| Act1 | Nuclear factor kappa B |
| ACKR3 | Atypical chemokine receptor 3 |
| ADAMTS-4 | A disintegrin and metalloprotease with thrombospondin motifs 4 |
| ADAMTS-5 | A disintegrin and metalloprotease with thrombospondin motifs 5 |
| AF | Annulus fibrosus |
| AGEs | Advanced glycation end-products |
| ALKBH5 | AlkB homolog 5 |
| ASK1 | Apoptosis signal-regulating kinase 1 |
| 5-AZA | 5-azacytidine |
| BCL10 | B cell lymphoma/leukemia 10 |
| CARMA3 | CARD-recruited membrane-associated protein 3 |
| cGAS–STING | Cyclic GMP-AMP synthase-stimulator of interferon genes |
| CCL2 | Chemokine (C-C motif) Ligand 2 |
| CCL3 | Chemokine (C-C motif) Ligand 3 |
| CCL4 | Chemokine (C-C motif) Ligand 4 |
| CCL17 | Chemokine (C-C motif) Ligand 17 |
| CCL20 | Chemokine (C-C motif) Ligand 20 |
| CCL22 | Chemokine (C-C motif) Ligand 22 |
| CCR2 | C-C Chemokine Receptor Type 2 |
| CEP | Cartilaginous endplates |
| COX2 | Cyclooxygenase-2 |
| COL2α1 | Collagen Type II Alpha 1 Chain |
| CRISPR/Cas9 | clustered regularly interspaced short palindromic repeats/CRISPR-associated protein 9 |
| CXCL1 | Chemokine (C-X-C motif) ligand 1 |
| CXCL10 | Chemokine (C-X-C motif) Ligand 10 |
| DAMPs | Damage associated molecular patterns |
| DDX3 | DEAD-box RNA helicase 3 |
| dECM | Decellularized extracellular matrix |
| DNMT | DNA methyltransferase |
| EBF1 | Early B-cell factor-1 |
| ECM | Extracellular matrix |
| EZH2 | Zeste homolog enhancer 2 |
| FTO | fat mass and obesity-associated protein |
| FCGR2A | Fc gamma receptor IIa |
| Gal3 | Galectin-3 |
| G-MDSCs | Granulocytic myeloid-derived suppressor cells |
| GSDMD | Gasdermin D |
| GSK3β | Glycogen synthase kinase 3 beta |
| HDAC | Histone deacetylase |
| HMGB1 | High mobility group box 1 protein |
| HP-sEVs | Hypoxia-preconditioned exosomal vesicles |
| IDD | Intervertebral disc degeneration |
| IKK | IκB kinase complex |
| IL-1β | Interleukin 1 beta |
| IL-6 | Interleukin 6 |
| IL-8 | Interleukin 8 |
| IL-10 | Interleukin 10 |
| IL-11 | Interleukin 11 |
| IL-16 | Interleukin 16 |
| IL-17 | Interleukin 17 |
| IFN-γ | Interferon-γ |
| IκBα | Inhibitor of Nuclear Factor Kappa B Alpha |
| IVD | Intervertebral disc |
| IRF7 | Interferon regulatory factor 7 |
| ITGAX | Identify integrin αX |
| JAK/STAT | Janus kinase/signal transducer and activator of the transcription |
| KLF5 | Krüppel-like factor 5 |
| LBP | Low back pain |
| LCN2 | Lipocalin-2 |
| LIF | Leukemia inhibitory factor |
| LPS | Lipopolysaccharides |
| MALT1 | Mucosa-associated lymphoid tissue lymphoma translocation protein 1 |
| MAPK | Mitogen-activated protein kinase |
| MARCHF8 | Membrane-associated ubiquitin ligase membrane-associated ring-CH-type finger 8 |
| MCP-1 | Monocyte chemoattractant protein-1 |
| M1-Exos | M1 macrophage-derived exosomes |
| M2-sEVs | M2 macrophage–derived small extracellular vesicles |
| METTL3 | Methyltransferase-like 3 |
| METTL14 | Methyltransferase-like 14 |
| MMP3 | Matrix metalloproteinase 3 |
| MMP9 | Matrix metalloproteinase 9 |
| MMP13 | Matrix metalloproteinase 13 |
| mtDNA | Mitochondrial DNA |
| MIF | Migration inhibitory factor |
| MSC | Mesenchymal stem cell |
| MR | Mendelian randomization |
| m6A | N6-methyladenosine |
| M2-sEVs | M2 macrophage–derived small extracellular vesicles |
| NF-κB | Nuclear factor kappa-B |
| NLRP3 | Nucleotide-binding oligomerization domain-like receptor protein3 |
| NOS2 | Nitric oxide synthase 2 |
| NP | Nucleus pulposus |
| NPC | Nucleus pulposus cell |
| NKT | Natural killer T |
| NK | Natural killer |
| NETs | Neutrophil extracellular traps |
| PGE2 | Prostaglandin E2 |
| PI3K–Akt | Phosphatidylinositol 3-kinase–protein kinase B |
| PINK1 | PTEN induced putative kinase1 |
| PPI | Protein–protein interaction |
| PPARγ | Peroxisome proliferator-activated receptor γ |
| PTEN | Phosphatase and tensin homolog deleted on chromosome ten |
| ROS | Reactive oxygen species |
| scRNA-seq | Single-cell RNA sequencing |
| SDC4 | Synergistically enhance syndecan-4 |
| SDF-1 | Stromal cell–derived factor-1 |
| SIRT3 | Sirtuin 3 |
| STAT3 | Signal Transducer and Activator of Transcription 3 |
| TNFR1/TNFR2 | Tumor necrosis factor receptor 1/2 |
| Th17 | T helper 17 |
| TAK1 | TGF-β–activated kinase 1 |
| TGFBI | Transforming growth factor-beta-induced protein |
| TLR | Toll like receptors |
| TNF-α | Tumor necrosis factor alpha |
| TGF-β | Transforming Growth Factor-beta |
| TRPV4 | Transient receptor potential vanilloid 4 |
| TSG-6 | TNF-stimulated gene 6 protein |